\documentclass[twocolumn,showpacs,preprintnumbers,amsmath,amssymb,prb]{revtex4}
\preprint{\hskip189pt LA-UR-02-5976}            

\usepackage{graphicx}
\usepackage{dcolumn}
\usepackage{bm}

\begin{document}


\title{Lattice dynamics and correlated atomic motion from the
atomic pair distribution function}

\author{I.-K. Jeong}
\altaffiliation {}
\email{jeong@lanl.gov}\homepage{http://www.totalscattering.org/}
\author{R. H. Heffner}
\author{M. J. Graf}
\affiliation{Los Alamos National Laboratory, Los Alamos,
NM 87545.}
\author{S. J. L. Billinge}%
\affiliation{%
Department of Physics and Astronomy and Center for Fundamental
Materials Research, Michigan State University, East Lansing, MI
48824-116
}%

\date{\today}

\begin{abstract}
The mean-square relative displacements (MSRD) of atomic pair
motions in crystals are studied as a function of pair distance and
temperature using the atomic pair distribution function (PDF). The
effects of the lattice vibrations on the PDF peak widths are
modelled using both a multi-parameter Born von-Karman (BvK) force
model and a single-parameter Debye model. These results are
compared to experimentally determined PDFs.  We find that the
near-neighbor atomic motions are strongly correlated, and that the
extent of this correlation depends both on the interatomic
interactions and crystal structure. These results suggest that
proper account of the lattice vibrational effects on the PDF peak
width is important in extracting information on static disorder in
a disordered system such as an alloy. Good agreement is obtained
between the BvK model calculations of PDF peak widths and the
experimentally determined peak widths. The Debye model
successfully explains the average, though not detailed, natures of
the MSRD of atomic pair motion with just one parameter. Also the
temperature dependence of the Debye model largely agrees with the
BvK model predictions. Therefore, the Debye model provides a
simple description of the effects of lattice vibrations on the PDF
peak widths.

\end{abstract}

\maketitle

\section{Introduction}

The pair distribution function (PDF) obtained from the powder
x-ray and neutron diffraction experiments has been shown to be of
great value in determining the local atomic structure of
materials.\cite{billi;bk98} The PDF results from a Fourier
transform of the powder diffraction spectrum (Bragg peaks +
diffuse scattering) into real-space.\cite{egami;b;lsfd98} For well
ordered crystals, apart from technical details, this is similar to
fitting the Bragg peaks + thermal diffuse scattering (TDS) in the
powder pattern in a manner first discussed by
Warren.\cite{warre;ac53}  A PDF spectrum consists of a series of
peaks, the positions of which give the distances of atom pairs in
real space. The ideal width of these peaks (aside from problems of
experimental resolution) is due both to relative thermal atomic
motion and to static disorder. Thus, an investigation of the
effects of lattice vibrations on PDF peak widths is important for
at least two reasons: first, to establish the degree to which
information on phonons (and the interatomic potential) can be
obtained from powder diffraction data, and, second, to account for
correlation effects in order to properly extract information on
static disorder in a disordered system such as an alloy.

In general, powder diffraction is not considered a favorable
approach for extracting information about phonons since, not only
is energy information lost in the measurement, but also the
diffuse scattering is isotropically averaged. The lattice
vibrations are best described from the phonon dispersion curves
determined using inelastic neutron scattering and
high-energy-resolution inelastic x-ray scattering on single
crystals.\cite{schwo;prl98,ruf;prl01} Nevertheless, with the
advent of high-energy synchrotron x-ray and pulsed-neutron sources
and fast computers, it is possible to measure data with
unprecedented statistics and accuracy. The PDF approach has been
shown to yield limited information about lattice vibrations in
powders,\cite{jeong;jpca99} though the extent of which this
information can be extracted remains
controversial.~\cite{dimit;prb99,reich;prb99,thorp;bk02,graf;02}

Measuring  powders has the benefit that the experiments are
straightforward and do not require single crystals. It is thus of
great interest to characterize the degree to which lattice
vibrations are reflected in the PDF using simple models, such as
the Debye model, in situations where detailed interatomic
potential information is not available. In this paper we explore
these issues by comparing both measured PDFs and those calculated
from realistic potential models with PDFs obtained through a
single-parameter Debye model. This comparison is carried out as a
function of atomic pair separation, temperature and direction in
the lattice. We find that a single parameter Debye model explains
much of the observed lattice vibrational effects on PDF peak
widths, including the temperature dependence, in crystals like Ni,
Ce, and GaAs. However, small but non-negligible deviations from
the Debye model calculation are evident in crystal which needs a
long-range interaction to explain anomalies in the dispersion
curves.\par

\section{correlated atomic motion in real-space}

\begin{figure}[!tb]
\includegraphics[angle=0,scale=0.4]{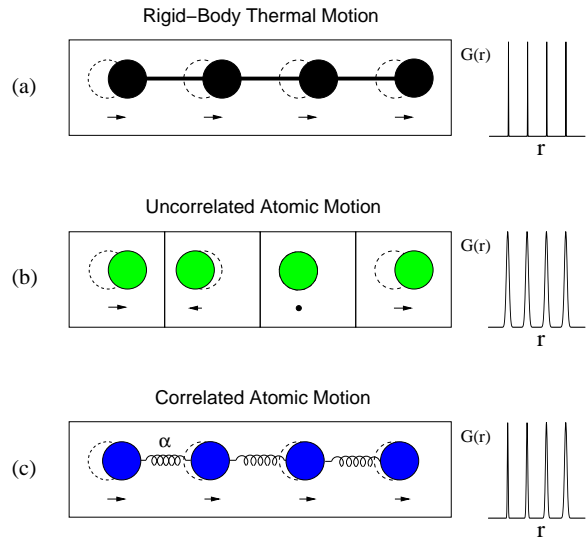}
\caption {Schematic diagram showing an instantaneous snapshot of
atomic positions in (a) rigid-body model (b) Einstein model (c)
Debye model. In (a) and (b) all PDF peaks have the same width
independent of atom separation. In (c) the PDF peak width
increases up to the root mean-square displacement as the atomic
separation increases. $\alpha$ is a spring constant.}
\label{fig;fig1}
\end{figure}

The existence of interatomic forces in crystals results in the
motion of atoms being correlated. This is usually treated
theoretically by transforming the problem to normal coordinates,
resulting in normal modes (phonons) that are non-interacting,
 thus making the problem mathematically tractible.
Projecting the phonons back into real-space coordinates yields a
picture of the dynamic correlations.  This situation can be
understood intuitively in the following way. Figure~\ref{fig;fig1}
shows a schematic diagram of atomic motion in three different
interatomic force systems, each with its corresponding ideal PDF
spectrum.  In a rigid-body system, Fig.~\ref{fig;fig1}(a), the
interatomic force is extremely strong and all atoms move in phase.
In this case,  the peaks in the PDF are delta-functions. At the
opposite extreme the atoms are non-interacting (the Einstein
model) and  move independently as shown in Fig.~\ref{fig;fig1}(b).
This type of atomic motion results in broad PDF peaks whose widths
are given by the root mean-square displacement amplitude ($\sqrt
{\langle u^2 \rangle}$).  In real materials, the interatomic
forces depend on atomic pair distances, i.~e., they are  strong
for nearest-neighbor interactions and get weaker as the atomic
pair distances increase. In fact, these interactions are often
quite well described with just nearest-neighbor or first- and
second-nearest-neighbor coupling. The case of nearest-neighbor
interactions is shown in Fig.~\ref{fig;fig1}(c). In this (Debye)
model a single parameter corresponding to the spring constant of
the nearest-neighbor interaction is used. Here, near-neighbor
atoms tend to move in phase with each other, while far-neighbors
move more independently. As a result, the near-neighbor PDF peaks
are sharper than those of far-neighbor pairs. This behavior was
first analyzed by Kaplow {\it et al.} in a series of
papers\cite{kaplo;jpcs64,kaplo;pr65,lagne;am67} for a number of
elemental metals.

\section{Experiments and Analysis}

The experimental PDFs discussed here were measured using pulsed
neutrons and synchrotron x-ray radiation. The neutron measurements
were carried out at the NPD diffractometer at the Manual Lujan,
jr., Neutron Scattering Center (LANSCE) at Los Alamos and the
x-ray experiments at beam line A2 at CHESS (Cornell). Powder
samples of Ni and a polycrystalline Ce rod were loaded into a
vanadium can for the neutron measurements, carried out at room
temperature. Powdered GaAs was placed between thin foils of kapton
tapes for the x-ray measurements, measured at 10~K using 60~KeV
($\lambda$ = 0.206~\AA) x-rays. Due to the higher x-ray energy at
CHESS and relatively low absorption coefficient of GaAs, symmetric
transmission geometry was used.

Both the neutron and x-ray data were
corrected~\cite{billi;prb93,jeong;prb01} for experimental effects
and normalized to obtain the total scattering function S(Q), using
programs PDFgetN~\cite{peter;jac00} and PDFgetX,\cite{jeong;jac01}
respectively. The experimental PDF, $G(r)$, was obtained by taking
the Fourier transform of  $S(Q)$: $G(r)={2 \over \pi}
\int_0^{Q_{max}} {\rm Q[S(Q)-1]sinQr\,dQ}$, where Q$_{max}$ is the
maximum momentum transfer. The experimental PDF peak widths as a
function of pair distance are extracted using the `real-space'
Rietveld program PDFFIT.~\cite{proffen;jac99} For detailed
procedures about modelling PDF spectrum and extracting PDF peak
widths refer to Ref. 6.

\section{Mean-Square Relative Displacements in Crystals}

\begin{figure}[!tb]
\includegraphics[angle=0,scale=0.8]{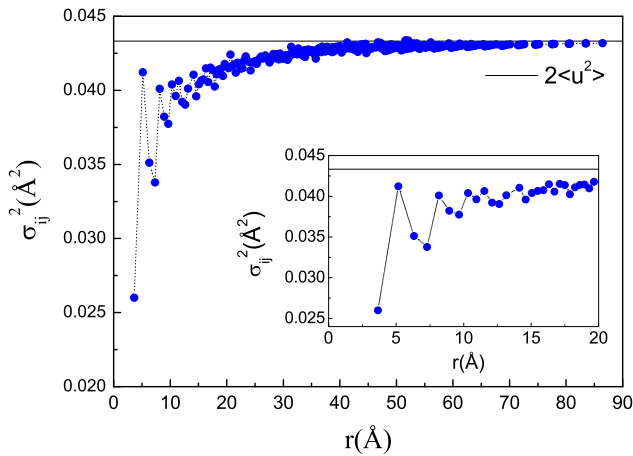}
\caption{Theoretical mean-square relative displacement
$\sigma^2_{ij}$ of $\gamma$-Ce as a function of pair distance
calculated using Eq.\ref{eq;msrd3} and the BvK model at 300 K. The
inset shows $\sigma^2_{ij}$ below r$_{ij} \leq$ 20 {\AA}. The
solid line corresponds to 2$\langle u^2 \rangle$, where $\langle
u^2 \rangle$ is the mean-square thermal displacement of
$\gamma$-Ce.} \label{fig;fig2}
\end{figure}

The PDF peak of simple materials can be well approximated by a
Gaussian function with a width
$\sigma_{ij}$.\cite{warre;bk90,chung;prb97} The mean-square
relative displacement of atom pairs, projected onto the vector
joining the atom pairs, is given by
\begin{equation}
\sigma^2_{ij} = \langle [({\bf u_i - u_j})\cdot {\bf \hat
r_{ij}}]^2 \rangle, \label{eq;msrd}
\end{equation}
where ${\bf u_i, u_j}$ are thermal displacements of atoms $i$ and
$j$ from their average positions.\cite{warre;bk90,chung;prb97} The
vector ${\bf \hat r_{ij}}$ is a unit vector parallel to the vector
connecting atoms $i, j$, and the angular brackets indicate an
ensemble average. This equation can be rearranged as
\begin{equation}
\sigma^2_{ij} = \langle [{\bf u_i \cdot \hat r_{ij}}]^2 \rangle
          + \langle [{\bf u_j \cdot \hat r_{ij}}]^2 \rangle
          -2\langle ({\bf u_i \cdot \hat r_{ij}})
              ({\bf u_j \cdot \hat r_{ij}}) \rangle .
\label{eq;msrd2}
\end{equation}
Here the first two terms correspond to mean-square thermal
displacement of atoms {\it i} and {\it j}.  The third term is a
displacement correlation function,  which carries information
about the motional correlations. For a monatomic crystal, the
$\sigma^2_{ij}$ is expressed in terms of the phonons as
follows,\cite{chung;prb97}
\begin{equation}
\sigma^2_{ij}
  = {2 \hbar \over NM}\,\sum_{{\bf k}, s} {{({\bf \hat e_{k,s}}
    \cdot {\bf \hat r_{ij}})^2 \over \omega_s({\bf k})} }
    [n(\omega_s({\bf k}))+{1 \over 2}][1-\cos(\bf {k \cdot r_{ij}})],
  \label{eq;msrd3}
\end{equation}
where $\omega_s({\bf k})$ is a phonon frequency with wave vector
${\bf k}$ in branch $s$, $n(\omega_s({\bf k}))$ is the phonon
occupation number, ${\bf \hat e_{k,s}}$ is the polarization vector
of the ${\bf k}, s$ phonon mode, $N$ is the number of atoms and
$M$ is the mass of an atom. As an example, we calculate the
$\sigma^2_{ij}$ of Ce (Fig.~\ref{fig;fig2}) using
Eq.~\ref{eq;msrd3}. Ce crystallizes in a simple FCC structure
(space group Fm3m) at room temperature and atmospheric
pressure.\cite{gschn;bk78}  In this calculation, the phonon
frequency ($\omega_s({\bf k})$) and polarization vector ($\bf
\hat{e}_{k,s}$) were obtained by solving the dynamical matrix
using up to the 8th nearest-neighbor (NN) interatomic force
parameters. The force parameters of Ce were determined by Stassis
{\it et al.}\cite{stass;prb79} by fitting the phonon dispersion
curves using the Born von-Karman (BvK) model.\cite{born;bk54} In
Fig.~\ref{fig;fig2} the horizontal solid line corresponds to
2$\langle u^2 \rangle$, where $\langle u^2 \rangle$ is the
mean-square thermal displacement of Ce.  Deviations from this line
are due to motional correlations in the pair motion. The inset
shows $\sigma_{ij}^2$ below $r_{ij} \leq$ 20 {\AA}, where the
motional correlation is more apparent.  Evidently, motion of
near-neighbor atoms is highly correlated, and this is reflected in
narrower PDF peak widths.  At larger separations ($r_{ij} \geq$
20~\AA) the $\sigma^2_{ij}$ values asymptotically approach  the
uncorrelated values because the cosine term in Eq.~\ref{eq;msrd3}
averages to zero.\par

As shown in Fig.~\ref{fig;fig2}, the motional correlation of atom
pairs varies significantly as a function of pair distance.
Therefore it is useful to quantify the degree of correlation using
a dimensionless correlation parameter $\phi$ which can be defined
as follows:\cite{booth;prb95,jeong;jpca99}
\begin{equation}
  \sigma_{ij}^{2}=\sigma_{i}^{2}+\sigma_{j}^{2}-2\sigma_{i}\sigma_{j}\phi,
  \label{eq;corrpara1}
\end{equation}
where $\sigma_{i}^2=\langle ({\bf u_i \cdot \hat r_{ij}})^2
\rangle$. It can be seen from Eq.~\ref{eq;corrpara1} that $\phi=0$
corresponds to completely uncorrelated motion. Positive values of
$\phi$ describe a situation where the atoms move in phase, and
thus the resulting value of $\sigma_{ij}$ is smaller than for the
uncorrelated case. Using Eq.~\ref{eq;corrpara1} the correlation
parameter $\phi$ can be calculated from the PDF peak width
measurements as
\begin{equation}
  \phi
  =\frac{(\sigma_i^{2}+\sigma_j^2)-\sigma_{ij}^{2}}{2\sigma_{i}\sigma_{j}}.
  \label{eq;corrpara2}
\end{equation}
\noindent
\begin{figure}[!tb]
\includegraphics[angle=0,scale=0.9]{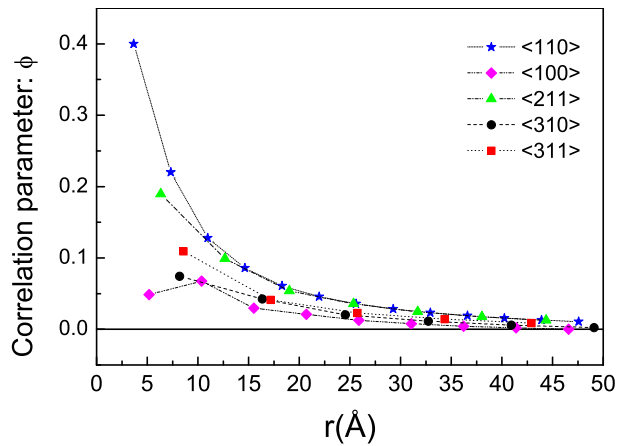}\vspace {0.0cm}
\caption{Correlation parameter $\phi$ of Ce atoms along various
crystallographic directions obtained using the BvK model
calculations at 300 K. $\phi$=0 corresponds to the uncorrelated
atomic motion and $\phi
>$ 0 indicates that atoms move in phase.}
\label{fig;fig3}
\end{figure}
Figure~\ref{fig;fig3} shows the motional correlations along
various crystallographic directions in Ce. It is clear that the
correlation parameter varies significantly with crystallographic
direction. Along the $<$110$>$ direction, Ce atomic motion shows
strong correlation. On the other hand, Ce atoms along $<$100$>$
show almost no motional correlation.  This behavior comes
principally from the characteristic elastic anisotropy of cubic
crystals.\cite{thorp;private} Despite the extensive orientational
averaging of the powder measurement, this directional information
survives in the data.

\begin{figure}[!tb]
\includegraphics[angle=0,scale=0.9]{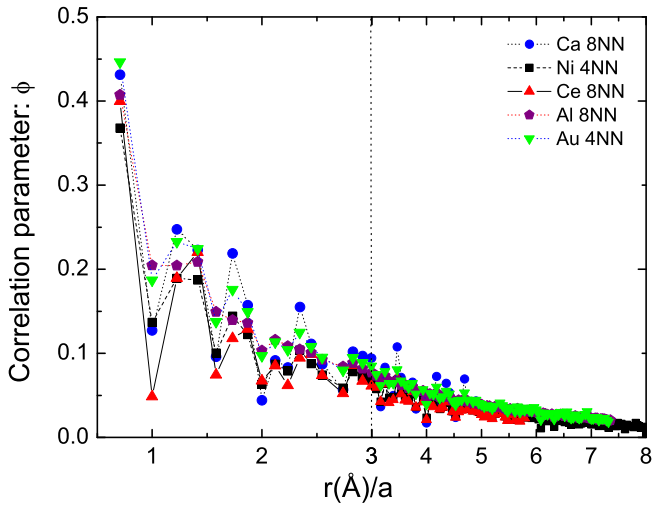}
\caption{Correlation parameter $\phi$ of FCC crystals Ca, Ni, Ce,
Al, Au at 300 K, obtained using the BvK model calculations.  {\bf
a} is the lattice parameter of each crystal.} \label{fig;fig4}
\end{figure}

\begin{figure}[!tb]
\includegraphics[angle=0,scale=0.9]{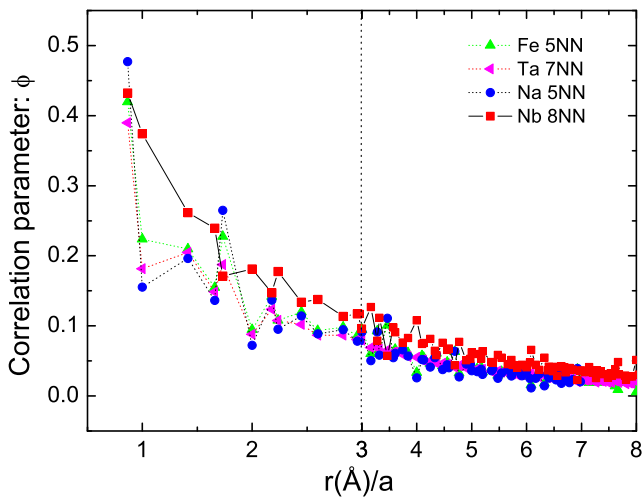}
\caption{Correlation parameter $\phi$ of BCC crystals Fe, Ta, Na,
Nb at 300 K, obtained using the BvK model calculations.  {\bf a}
is the lattice parameter of each crystal.} \label{fig;fig5}
\end{figure}

The oscillations in $\sigma^2_{ij}$ shown in Fig.~\ref{fig;fig2}
for Ce are generally driven by both the interatomic interactions
and the crystal structure. This is illustrated in
Figs.~\ref{fig;fig4} and~\ref{fig;fig5}, which show the
correlation parameter $\phi$ for a variety of FCC and BCC
materials. The $\sigma^2_{ij}$ in Figs.~\ref{fig;fig4}
and~\ref{fig;fig5} are generated in the same way as for Ce, that
is, using the BvK force model, with parameters derived from fits
to the phonon dispersion curves found in the
literature.\cite{landolt;vol13}  One sees that a common
oscillatory behavior in the correlation parameter is found for all
of the FCC crystals studied.  And similar but distinct oscillatory
features are observed for the BCC crystals, except Nb. This
difference in the general behavior of FCC and BCC crystals
suggests that atomic geometry plays a role in the motional
correlations as well as the interatomic interaction. However,
significant differences in the correlation are evident among
different elements with the same crystal structure as well. For
example, in FCC crystals $\phi$ varies from 0.37 to 0.45 for the
1NN pair and from 0.05 to 0.2 for the 2NN pair. In BCC crystals
$\phi$ varies from 0.38 to 0.47 for the 1NN pair. Thus, $\phi$ and
$\sigma^2_{ij}$  do reflect interatomic interactions.\par
\noindent
\noindent

\begin{figure}[!tb]
\includegraphics[angle=0,scale=1.0]{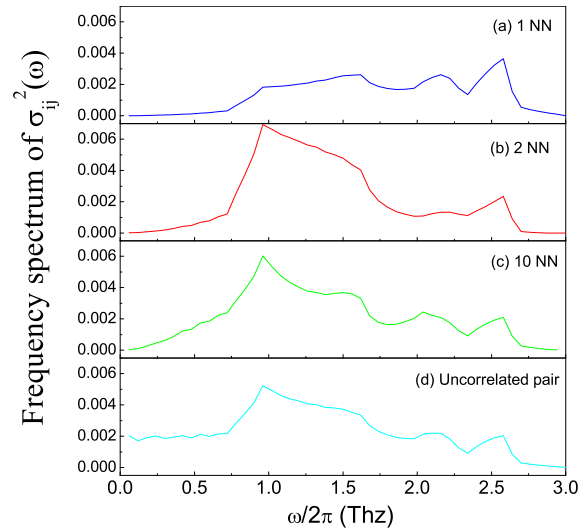}
\caption{Frequency spectrum of $\sigma_{ij}^2$ for Ce at 300 K
obtained using the BvK model calculations: (a) 1NN, (b) 2NN, (c)
10NN, and (d) uncorrelated far-neighbor atom pair. The area
underneath the solid line in each figure corresponds to the
$\sigma_{ij}^2$ of each pair.} \label{fig;fig6}
\end{figure}

In addition to a dependence on the atom pair distance $r_{ij}$,
the $\sigma^2_{ij}$ shows an explicit dependence on phonon
frequencies $\omega_k(s)$. Therefore, it is instructive to
consider how phonon modes of different frequencies contribute to
the broadening of the peak widths. Figure~\ref{fig;fig6} shows the
frequency spectrum of $\sigma^2_{ij}$ in Ce for the 1NN, 2NN and
10NN peaks, in addition to that of the uncorrelated far-neighbor
atom pairs. Referring to Fig.~\ref{fig;fig6}, for the 1NN, the low
frequency ($\omega/2\pi \leq$~1~THz) thermal motion contributes
little to the 1NN PDF peak width broadening. Most of the peak
broadening comes from mid-to-high frequency modes (1 $\leq
\omega/2\pi \leq$ 3 THz), which contribute almost equally to the
broadening.  This suggests that 1NN pair moves more or less
in-phase at low frequencies, and that the pair motion de-phases
somewhat as the frequency increases. For the 2NN and
higher-neighbors peaks, however, where the motion of atom pairs
gets more and more de-phased as the pair distance increases, the
medium-range frequency modes predominantly contribute to peak
width broadening. Finally, for the far-neighbor atom pairs, where
the motion is completely uncorrelated, the frequency spectrum is
spread more evenly across all of the frequencies.\par

\section{Correlated Thermal Motion: Debye approximation}

Using the BvK model we have shown that near-neighbor atomic
motions in crystalline materials are strongly correlated. In the
BvK model calculation, however, the force constants must be known
in advance. In this section, we simplify the result in
Eq.~\ref{eq;msrd3} using some approximations to describe the
effects of the lattice vibrations on the PDF peak widths without
knowing the force constants. Following Debye\cite{debye;ap12} and
Beni and Platzmann,\cite{beni;prb76} we make no distinction
between longitudinal and transverse phonon modes and take a
spherical average. Then Eq.~\ref{eq;msrd3} reduces to
\begin{equation}
\sigma^2_{ij}
  = \biggl < {2 \hbar \over M \omega}
     \Bigl [n(\omega)+{1\over 2}\Bigr]\Bigl[1-\cos(\bf {k \cdot
r_{ij}})\Bigr] \biggr >,
  \label{eq;msrdave}
\end{equation}
where $\langle \cdot \cdot \cdot \rangle$ is the average over the
3$N$ modes and $N$ is the number of atoms. This result is a
general expression for all materials and is independent of the
number of atoms per unit cell.\cite{thorp;private} Using the Debye
approximation, $\omega = ck$, we can write Eq.~\ref{eq;msrdave} as
follows~\cite{sevil;prb79}:
\begin{equation}
\sigma^2_{ij}
  = {2 \hbar \over 3NM}\,\int_{0}^{\omega_D}  d\omega \,{\rho (\omega) \over
\omega} \biggl [n(\omega)+{1\over 2} \biggr]\biggl[1-{\sin(\omega
r_{ij}/c) \over \omega r_{ij}/c} \biggr],
  \label{eq;msrddw}
\end{equation}
where $\rho (\omega) = 3N (3 \omega^2/ {\omega_D}^3) $ is the
phonon density of states, n($\omega$) is the phonon occupation
number, $c$ is the sound velocity and $\omega_D ~{= c k_D}$ is the
Debye cut-off frequency. The Debye wavevector is given by $k_D =
(6\pi^2N/V)^{1/3}$ where $N/V$ is the number density of the
crystal. After integrating over $\omega$, we obtain
\begin{eqnarray}
\sigma^2_{ij} &=& {6 \hbar \over M
\omega_D}\biggl[\frac{1}{4}+\biggl(\frac{T}{\Theta_D}\biggr)^2
\Phi_1 \biggr] - {6 \hbar \over M \omega_D} \biggl[ {1-\cos(k_D
r_{ij}) \over {2(k_D r_{ij})^2}}
\nonumber \\
&+& \biggl(\frac{T}{\Theta_D}\biggr)^2 \int_{0}^{\frac{\Theta_D}
T} \frac{\sin(\frac{k_D r_{ij} T x}{\Theta_D})/(\frac{k_D r_{ij}
T}{\Theta_D})} {e^x-1}~dx \biggr], \label{eq;msrd4}
\end{eqnarray}
where $\Phi_1 = \int_{0}^{\Theta_D/T} x (e^x-1)^{-1}~dx$, $x$ is a
dimensionless integration variable and $\Theta_D$ (=$\hbar
\omega_D/k_B$) is the Debye temperature. This result is known as
the ``correlated Debye (CD)
model''.\cite{beni;prb76,bohme;jpcs79,sevil;prb79} Here, the first
term corresponds to the usual uncorrelated mean-square thermal
displacements (2$\langle u^2 \rangle$) and the second term is the
displacement correlation function (DCF). The CD model was first
used to explain XAFS peak widths as a function of temperature, and
provided reasonable fits to the 1NN and 2NN peak widths. However,
this model has never been tested beyond the 2NN peak. Here we test
this simple model against full experimental PDF spectra and the
BvK model calculations.\par
\begin{figure}[h]
\includegraphics[angle=0,scale=1.0]{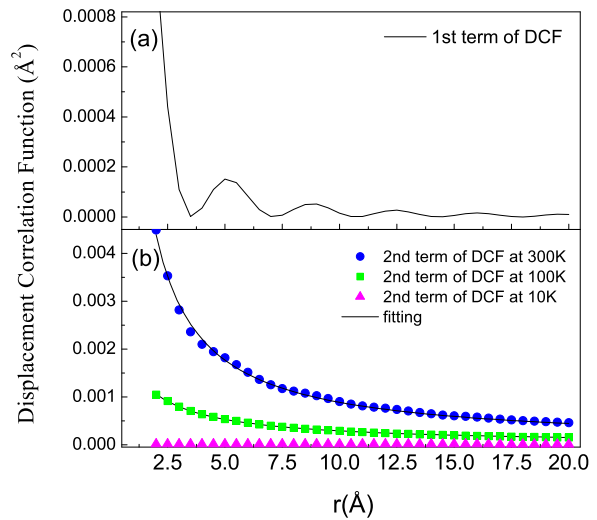}
\caption{ Displacement Correlation Function (DCF) of Ni. (a) 1st
term of DCF of Eq.~\ref{eq;msrd4}.
  (b) second term of DCF of Eq.~\ref{eq;msrd4}
  at 300~K, 100~K, and 10~K. Solid line is a fit to a  $1/r$ dependence.}
  \label{fig;fig7}
\end{figure}

In the CD model the DCF shows explicit dependence on the atomic
pair distance $r_{ij}$. The first term of the DCF comes from the
quantum zero-point motion and the second term is temperature
dependent. Figure~\ref{fig;fig7} shows the $r_{ij}$ dependence of
the first and second terms of the DCF of Ni. The first term of the
DCF decreases as $1/r_{ij}^2$ with a cosine modulation. The second
term is temperature dependent and shows a $1/r_{ij}$ dependence.
At low temperatures (T$\ll \theta_D$), the temperature dependent
DCF term is much smaller than the zero-point motion term. However,
as the temperature increases the second term becomes dominant.
These results show that the motional correlation follows a
$1/r_{ij}^2$ dependence when T$\ll \theta_D$ and a $1/r_{ij}$ when
T$\geq \theta_D$.\par

\begin{figure}[h]
\includegraphics[angle=0,scale=1.0]{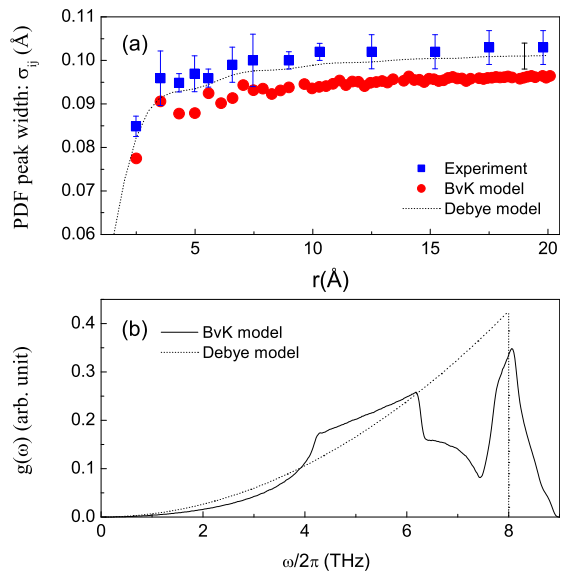}
\caption{(a) Comparison of neutron PDF and BvK model peak widths
with those of the CD model calculation of Ni at 300~K. Filled
squares: experimental PDF peak widths (only a few selected peak
widths are shown), Filled circles: BvK model calculations, Dotted
line: CD model calculation using Debye temperature
$\theta_D$=385~K, Debye wavevector $k_D$=1.756 \AA$^{-1}$. (b)
Solid line: Ni phonon density of states calculated using BvK
model. Dotted line: Debye density of states with the same area as
the BvK calculation. Debye cut-off frequency $\omega_D/2\pi$=8
THz.} \vspace {0.0cm} \label{fig;fig8}
\end{figure}

We tested the CD model calculation against the experimental and
BvK PDF peak widths of Ni at 300 K. Figure~\ref{fig;fig8} shows
selected experimental PDF peak widths and the calculated peak
widths as a function of pair distance, as well as the
corresponding phonon density of states of Ni. The errors in the
experimental PDF peak widths were estimated from fitting Gaussian
functions to the PDF peaks. The error in CD model calculation was
estimated from the error in the experimental thermal displacement.
For the BvK model calculation, we used up to the 4th NN force
parameters determined by Birgeneau {\it et al.}.\cite{birge;pr64}
We also compared the 4NN BvK model calculations with those of a
simple 1NN BvK model and found excellent agreement between them.
Our experimental thermal displacement of Ni ($\langle u^2 \rangle$
= 0.00535 \AA$^2$) is about 10\% larger than that of the BvK model
calculation. As a result, as shown in Fig. \ref{fig;fig8}(a), the
BvK model peak widths are shifted downward by roughly 5 \%
overall. The origin of this difference between our thermal
displacement and that of the BvK calculation is not clear but the
Debye temperature, determined from our thermal displacement using
the Debye approximation, $\theta_D$=385~K, compares well to the
specific heat measurement, where
$\theta_D$=375~K.~\cite{achcr;i;bk76} For the CD model calculation
the Debye wavevector $\rm k_D$=1.756~{\AA$^{-1}$} was obtained
from the atomic geometry. The only parameter in the CD model, the
Debye temperature, was determined from the experimental thermal
displacement.

As shown in  Fig. \ref{fig;fig8}(b), the Debye model reasonably
approximates the `real' density of states in this simple element.
However, because the Debye temperature and corresponding Debye
frequency were obtained from the thermal displacement, and not
from the sound velocity, the low-frequency Debye density of states
deviates from those of the BvK model by roughly 15\%.
Nevertheless, the CD model calculation of the peak widths shows
good agreement with the experimental PDF peak widths $\sigma_{ij}$
except for the overall downward shift.

\begin{figure}[h]
\includegraphics[angle=0,scale=1.0]{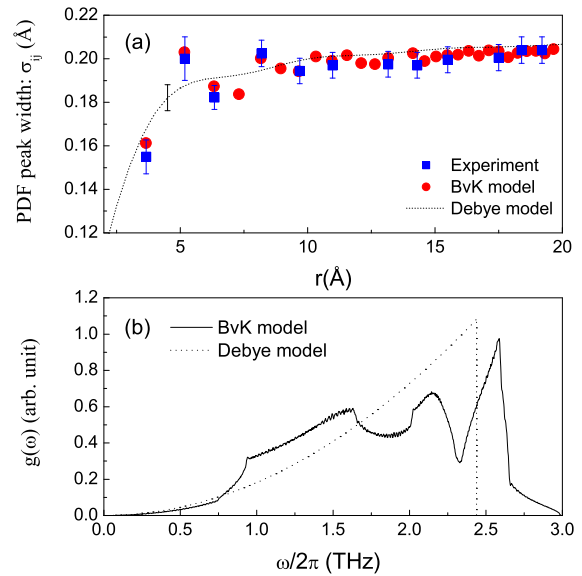}
\caption{(a) Comparison of neutron PDF and BvK model peak widths
with those of the CD model calculation of Ce at 300~K. Filled
squares: experimental PDF peak widths (only a few selected peak
widths are shown), Filled circles: BvK model calculation, Dotted
line: CD model calculation with Debye temperature
$\theta_D$=117~K, Debye wavevector $k_D$=1.1986 \AA$^{-1}$. (b)
Solid line:  Ce phonon density of state calculated using BvK
model. Dotted line: corresponding Debye density of states, using
Debye cut-off frequency $\omega_D/2\pi$=2.44 THz.}
\label{fig;fig9}
\end{figure}

We now move to the more complex case of Ce, which needs long-range
forces to explain anomalies in the dispersion curves. Up to the
8th NN interactions are required to reasonably fit the phonon
dispersion curves of Ce.\cite{stass;prb79} Figure~\ref{fig;fig9}
shows selected experimental PDF peak widths and calculated peak
widths as a function of pair distance, as well as the phonon
density of states of Ce. In this case, the BvK calculation of the
peak widths shows a good agreement with the experimental PDF peak
widths. We also found that a simple 1NN BvK model calculation
shows very good overall agreement with that of 8NN BvK model,
except for the 2NN PDF peak width which deviates by $\sim$ 3 \%.
For the CD model calculation, a Debye wavevector $k_D$=1.1986
\AA$^{-1}$ and a Debye temperature $\theta_D$=117~K (at 300~K)
were obtained from the atomic geometry and thermal displacement of
Ce ($\langle u^2 \rangle $=0.0231 \AA$^2$), respectively. Even for
the more complex system like Ce, the CD model calculation of the
PDF peak widths reproduces the overall $r_{ij}$-dependence rather
well, except for a few detailed features.\par

\begin{figure}[h]
\includegraphics[angle=0,scale=1.0]{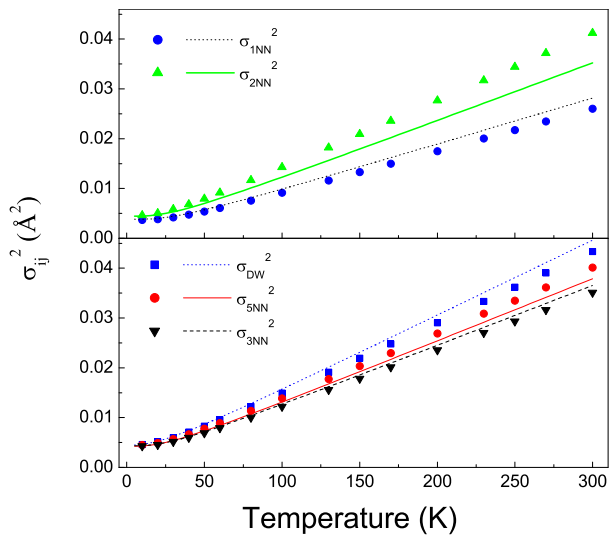}
\caption{Mean-square relative displacement $\sigma^2_{ij}$ of Ce
as a function of temperature. Upper panel: nearest neighbor (NN),
second NN (2NN). Lower panel: third NN (3NN), fifth NN (5NN) and
DW. DW represents uncorrelated far-neighbor pair. Symbols are the
Born von-Karman (BvK) model calculations and lines are the
corresponding CD model calculations. In BvK calculation, the Debye
temperature, $\theta_D$=117~K is determined at
300~K.}\label{fig;fig10}
\end{figure}

We now consider the $T$-dependence of  $\sigma^2_{ij}$ for
specific atom pairs of Ce. These are shown in Fig.~\ref{fig;fig10}
for the first few NN pairs, as well as the uncorrelated atom pair.
All of these curves exhibit characteristic Debye-like behavior,
i.e., linearity in $T$ at high temperatures, but curving over to a
common zero-point value at $T=0$. In general, the CD model
calculation of the temperature dependence of the $\sigma^2_{ij}$
shows good agreement with that of the BvK model calculation,
except for the 2NN pair (Fig.~\ref{fig;fig10}). Referring to
Fig.~\ref{fig;fig9} we see that BvK calculations of the
$\sigma^2_{ij}$ of the 1NN and 3NN lie very close to the Debye
prediction, but the 2NN is displaced significantly upwards. The
deviations between the BvK and Debye models increase as the
temperature increases. Apparently, if the $\sigma^2_{ij}$ lies on
the Debye prediction at one temperature, the Debye model will also
predict its temperature dependence correctly. Conversely, the
temperature dependence of $\sigma^2_{ij}$ will be underestimated
or overestimated depending on whether it is displaced above or
below the Debye curve at the lowest temperatures, respectively.
This is at least true for Ce.\par

\begin{figure}[h]
\includegraphics[angle=0,scale=1.0]{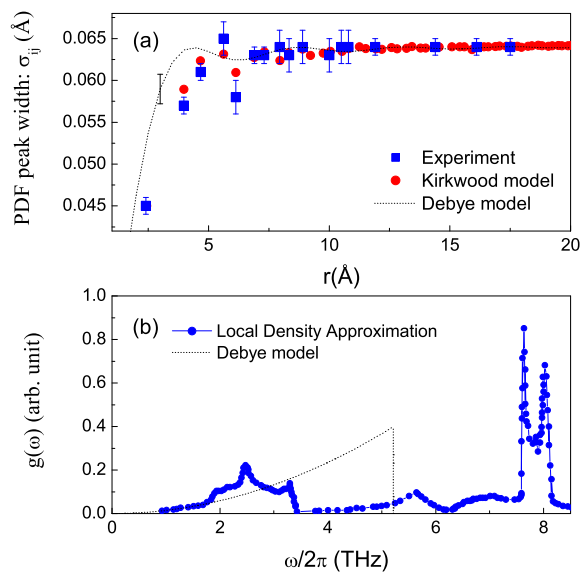}
\caption{(a) PDF peak widths of GaAs as a function of atom
separation at 10 K. Filled squares: Experimental x-ray PDF peak
widths, Filled circles: lattice dynamics calculation using the
Kirkwood model ($\alpha$=96~N/m, $\beta$=10~N/m), Dotted line:  CD
model calculation using the Debye temperature $\theta_D$=250~K,
Debye wavevector $k_D$=1.382 \AA$^{-1}$. (b) Symbols: GaAs phonon
density of states calculated using local-density approximation
density functional theory~\cite{pavone;thesis}, Solid line: Debye
density of states with the same area as the LDA calculation. Debye
cut-off frequency $\rm \omega_D/2\pi=5.22$ THz.} \label{fig;fig11}
\end{figure}

As a final example, we compare the CD model calculation of PDF
peak widths with those of GaAs determined experimentally. This is
distinct from the above examples due to the presence both of more
than one atomic species and directional covalent bonding. We also
compare the CD model calculation with a lattice-dynamics
calculation using the Kirkwood potential, which has been shown to
be a good basis for describing semiconductor compounds and
alloys.\cite{petko;prl99,jeong;prb01} Figure~\ref{fig;fig11}(a)
shows the atom pair dependence of the PDF peak widths of GaAs at
10~K. In the Kirkwood model, the potential parameters are obtained
by fitting the nearest and far neighbor PDF peak
widths.\cite{petko;prl99, jeong;prb01} In Fig.~\ref{fig;fig11}(a),
the Kirkwood model calculation using bond stretching
($\alpha$=96~N/m) and bending ($\beta$=10~N/m) force constants
shows quite good agreement with the experimental PDF peak widths.
In the CD model, the average mass of Ga and As is used. The Debye
wavevector $k_D=1.382$ {\AA$^{-1}$} and Debye temperature
$\theta_D$=250~K were obtained from the atomic geometry and by
fitting the far-neighbor PDF peak widths. In the CD model
calculation, the near-neighbor peaks ($r \leq 5$ \AA) are $\sim$
10\% broader than those of the experimental peaks. Referring to
Fig.~\ref{fig;fig11}(b), the Debye model does a poor job of
describing the GaAs phonon density of states; for example, the
high frequency optic modes, 6 $\leq \omega/2\pi \leq $8~THz, are
totally missed. Instead, the Debye model over-estimates phonon
modes between 3.5 THz $\leq \omega/2\pi \leq  \omega_D$. This poor
description of the phonon density of states, as well as the use of
the average mass of Ga and As for $M$ in Eq.~\ref{eq;msrddw},
leads to deviations in near-neighbor peak widths from those of the
experimental peaks and the BvK model calculations. Therefore, the
deviations in CD model calculations basically reflect the
limitation of the CD model in describing motional correlations in
a system with optic modes. Nevertheless, the CD model, with a
single parameter $\theta_D$ determined from the thermal
displacement, serves as a good first order approximation to the
PDF peak widths, even in more complex systems like GaAs.\par.

\section{Discussion}

The mean-square relative displacements $\sigma^2_{ij}$ and the
corresponding correlation parameter shown in Figs.~\ref{fig;fig2},
\ref{fig;fig4}, \ref{fig;fig5} and \ref{fig;fig8}, \ref{fig;fig9}
present two interesting pieces of information about the atomic
motions in a crystalline material. First of all, they show that
nearest-neighbor atomic motion is significantly correlated.
Second, the details of the motional correlations as a function of
pair distance display structures which deviate from the
predictions of the simple CD model. Here we can raise some
interesting questions. How is this structure in the motional
correlation of atom pairs related to the underlying interatomic
potentials? Can one extract the potential parameters using an
inverse approach to model the PDF peak widths with the potential
parameters as input?

Reichardt and Pintschovius\cite{reich;prb99} argued that the
calculated PDF peak widths as a function of pair distance are
rather insensitive to the details of the lattice dynamics models
used to calculate $\sigma^2_{ij}$.  They found that PDFs
calculated using either very simple or complex models didn't show
significant differences.  A similar conclusion has been reached by
Graf {\it et al.},~\cite{graf;02} in contradiction to previous
claims by Dimitrov {\it et al.}.~\cite{dimit;prb99}  Indeed, the
magnitude of errors implicit in the measurement and data analysis
appear to be comparable to the effects that must be measured to
obtain quantitatively accurate potential information using this
approach.~\cite{thorp;bk02} The conclusions of Reichardt and
Pintschovius and Graf {\it et al.} and Thorpe {\it et al.} are
largely borne out by the present work; e.g., the grossly
oversimplified CD model, which neglects elastic anisotropy and
parameterizes the dynamics with a single number $\theta_D$, is
rather successful at explaining the smooth $r_{ij}$-dependence of
the PDF peak widths.\par

Thus, when the BvK force parameters are not available, we have
shown that the correlated Debye (CD) model is a reasonable
approximation to describe both the smooth $r_{ij}$-dependence and
the temperature dependence of $\sigma_{ij}^2$ in simple elements.
Considering the poor correspondence between the Debye phonon
density of states and the BvK density of states, the reasonable
agreement between the BvK model calculations of $\sigma^2_{ij}$
and that of the CD model is rather surprising. This confirms that
the PDF peak width is rather insensitive to the details of the
phonon density of states and the phonon dispersion curves, as
suggested by Reichardt and Pintschovius and by Graf {\it et al.}.
Any information about the interatomic forces in the PDF peak
widths is contained in the small deviations of the $\sigma^2_{ij}$
from those of the CD model calculations. Therefore, extracting
interatomic potential information from the PDF peak widths is
unlikely. However, these deviations could possibly yield some
average phonon information. For example, recent calculations by
Graf {\it et al.}\cite{graf;02} show that one can obtain phonon
moments within a few percent accuracy for most FCC and BCC
crystals using the nearest-neighbor force parameters extracted
from a theoretical BvK PDF spectrum. This result indicates that
the PDF spectrum contains some average phonon information,
although it doesn't provide detailed phonon dispersion
information. The average phonon information, such as phonon
moments from the PDF peak widths, will be a useful complement to
optical and acoustic techniques that yield zone-center information
in situations where single crystal measurements are not possible.
This complementarity also extends to the extraction of
Debye-Waller factors from powder diffraction measurements.

Finally, a comparison of the CD model calculations of the PDF peak
widths in GaAs with those of experimental PDF and Kirkwood model
calculations shows additional limitations of the CD model. In the
CD model calculation, the near-neighbor PDF peaks below $r\leq$
5{~\AA} are about 5-10\% broader than those of experimental PDF
peaks. This is due to the poor description of GaAs phonon density
of states by the Debye model. Since the sine term in
Eq.~\ref{eq;msrddw} over- and under-weighs certain phonon modes
depending on their frequencies, the re-distribution of GaAs phonon
density of states in a realistic model causes deviations in
near-neighbor peak widths from those of the CD model. One way to
improve the model calculation in materials which have optic modes
might be a hybrid model that combines the correlated Debye and
Einstein models. Such a hybrid model has worked quite well in the
case of AgI.\cite{dalba;prb90}

\section{Summary}

In this paper the mean-square relative displacements
($\sigma_{ij}^2$) of atomic pair motion in crystals have been
studied as a function of pair distance and temperature using the
atomic pair distribution function (PDF). The experimental PDF peak
width and the BvK model calculations of $\sigma^2_{ij}$ as a
function of pair distance show that the near-neighbor atomic
motions are strongly correlated. The extent of these correlations
depends both on the interatomic interactions and crystal
structure. Thus, a proper accounting of the lattice vibrational
effects on the PDF peak widths is important in order to better
understand the effects of static and dynamic disorder on the PDF
peak widths in disordered systems. Details of the PDF peak widths
vs. $r_{ij}$ seen in the BvK calculations are well reproduced in
the measured data indicating the accuracy of the measurements.
Most of these details originate from the elastic anisotropy of the
crystal which is especially apparent in FCC crystals. We showed
that the CD model reproduces the average features of the lattice
vibrational effects on the PDF peak width with just one parameter,
which is determined from the measured thermal displacement
$\langle u^2 \rangle$. Therefore, this simple model can be used as
an important adjunct when using PDF to extract static and dynamic
disorder disorder information from materials with local lattice
distortion. In addition, the {\it T}-dependence of the CD model
largely agrees with the BvK model calculations. Good agreement
between CD model and experimental PDF peak widths indicates that
the PDF peak widths are rather insensitive to the details of
phonon density of states and the phonon dispersion curves. Any
information about the interatomic forces in the PDF peak widths is
contained in the small deviations ($\leq$ 5\%) of the
$\sigma_{ij}$ from those of the CD model calculation. This makes
the extraction of interatomic potential information from PDF peak
widths alone unlikely.
\par
\acknowledgements

IKJ gratefully acknowledges Prof. M. F. Thorpe for many helpful
discussions. We thank Drs. T. Darling, Th. Proffen and A. C.
Lawson for the help with the data collections. SJLB acknowledges
support from US Department of Energy, Office of Science, through
grant DE-FG02-97ER45651. Works by IKJ, MJG and RHH were carried
out under the auspices of the US Department of Energy, Office of
Science. Part of the data were collected at the Manuel Lujan, Jr.
Neutron Scattering Center which is a national user facility funded
by the U.S. Department of Energy, Office of Basic Energy Sciences.
And part of the work is based upon research conducted at the
Cornell High Energy Synchrotron Source (CHESS) which is supported
by the National Science Foundation and the National Institutes of
Health/National Institute of General Medical Sciences under award
DMR 9713424.


\end{document}